\documentclass{PoS}
\usepackage{verbatim}
\usepackage{epsfig}

\newcommand{\fatlink}{\mbox{\raisebox{-0.15mm}
{\epsfig{file=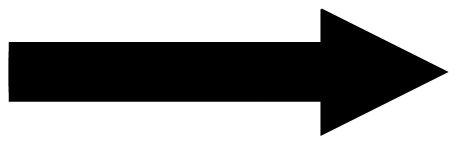,height=1.1mm
}}~}}
\newcommand{\fatlinka}{\mbox{\raisebox{-0.15mm}
{\epsfig{file=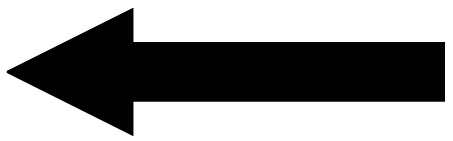,height=1.1mm
}}~}}
\newcommand{\alink}{\mbox{
\begin{picture}(2.5,.2)
\linethickness{1mm}
\multiput(0,0.1)(2,0){2}{\circle*{0.1}}
\multiput(1.01,0.1)(0,0){1}{\circle{0.2}}
\put(1,0.){\fatlink}
\put(0,0.){\fatlinka}
\put(1,-.5){\scriptsize \( i \) }
\put(-.5,-.5){\scriptsize \( j \) }
\put(1.9,-.5){\scriptsize \( j \) }
\end{picture}}}
\newcommand{\blinkba}{\mbox{
\begin{picture}(2.5,2.5)
\thicklines
\multiput(1,-1)(0,-1){2}{\circle*{0.1}}
\multiput(1,0)(0,0){1}{\circle{0.2}}
\multiput(2,0.0)(0,1){3}{\circle*{0.1}}
\multiput(0,-2)(0,0){1}{\circle*{0.1}}
\put(1,-2.0){\vector(-1,0){1}}
\put(1,-1.0){\vector(0,-1){1}}
\put(1,0.0){\vector(0,-1){1}}
\put(1,0.0){\vector(1,0){1}}
\put(2,0.0){\vector(0,1){1}}
\put(2,1.0){\vector(0,1){1}}
\put(0.5,-0.1){\scriptsize \( i \) }
\put(1.5,2.0){\scriptsize \( j \) }
\put(-0.4,-2.1){\scriptsize \( j \) }
\end{picture}}}
\newcommand{\blinkbb}{\mbox{
\begin{picture}(2.5,2.5)
\thicklines
\multiput(0,0.0)(0,-1){3}{\circle*{0.1}}
\multiput(1,1.0)(0,1){2}{\circle*{0.1}}
\multiput(1,0.0)(0,){1}{\circle{0.2}}
\multiput(2,2)(0,1){1}{\circle*{0.1}}
\put(0,-1.0){\vector(0,-1){1}}
\put(0,0){\vector(0,-1){1}}
\put(1,0){\vector(-1,0){1}}
\put(1,0){\vector(0,1){1}}
\put(1,1){\vector(0,1){1}}
\put(1,2){\vector(1,0){1}}
\put(1.25,-0.1){\scriptsize \( i \) }
\put(2.2,2.0){\scriptsize \( j \) }
\put(-0.4,-2.1){\scriptsize \( j \) }
\end{picture}}}
\newcommand{\blinkbc}{\mbox{
\begin{picture}(2.5,2.5)
\thicklines
\multiput(1,1)(0,1){2}{\circle*{0.1}}
\multiput(1,0)(0,0){1}{\circle{0.2}}
\multiput(2,0)(0,-1){3}{\circle*{0.1}}
\multiput(0,2)(0,1){1}{\circle*{0.1}}
\put(1,2){\vector(-1,0){1}}
\put(1,1){\vector(0,1){1}}
\put(1,0){\vector(0,1){1}}
\put(1,0){\vector(1,0){1}}
\put(2,0){\vector(0,-1){1}}
\put(2,-1){\vector(0,-1){1}}
\put(0.5,-0.1){\scriptsize \( i \) }
\put(1.5,-2.1){\scriptsize \( j \) }
\put(-0.5,2.0){\scriptsize \( j \) }
\end{picture}}}

\newcommand{\blinkbd}{\mbox{
\begin{picture}(2.5,2.5)
\thicklines
\multiput(0,0)(0,1){3}{\circle*{0.1}}
\multiput(1,-1)(0,-1){2}{\circle*{0.1}}
\multiput(1,0)(0,-1){1}{\circle{0.2}}
\multiput(2,-2)(0,1){1}{\circle*{0.1}}
\put(0,1){\vector(0,1){1}}
\put(0,0){\vector(0,1){1}}
\put(1,0){\vector(-1,0){1}}
\put(1,0){\vector(0,-1){1}}
\put(1,-1){\vector(0,-1){1}}
\put(1,-2){\vector(1,0){1}}
\put(1.25,-0.1){\scriptsize \( i \) }
\put(2.1,-2.1){\scriptsize \( j \) }
\put(-0.5,2.0){\scriptsize \( j \) }
\end{picture}}}

\long\def\symbolfootnote[#1]#2{\begingroup%
\def\thefootnote{\fnsymbol{footnote}}\footnote[#1]{#2}\endgroup}

\title{Analysis of the finite temperature transition for 3 flavor QCD using p4-improved staggered fermions}

\ShortTitle{Analysis of the finite temperature transition for 3 flavor QCD using p4-improved staggered fermions}

\author{\speaker{Michael Cheng} for the RBC-Bielefeld Collaboration\\
Department of Physics, Columbia University, New York, NY 10027, USA\\
E-mail: \email{michaelc@phys.columbia.edu}
}

\abstract{
We present a calculation of the transition temperature for 3 flavor QCD
using p4-improved staggered fermions with two different variants of fattened
links. We examine various susceptibilities at two different values for the
temporal extent - $N_t = 4$ and $N_t = 6$ - in the vicinity of the 3
flavor transition at vanishing chemical potential.
For $N_t = 4$, we study bare quark masses in the range $m_qa = 0.005$
to $m_qa = 0.1$, and for $N_t = 6$, we use $m_qa = 0.02$ to $m_qa =
0.2$. We also calculate hadron masses and the static quark potential at zero
temperature in order to set the scale for the transition temperature
and to study quark mass and cut-off dependent scaling properties of the
transition temperature.  A comparison of the R Algorithm and the RHMC 
algorithm for finite-temperature simulations is also made.}

\FullConference{XXIVth International Symposium on Lattice Field Theory\\
                July 23-28, 2006\\
                Tucson, Arizona, USA}

\begin{document}

\section{Introduction}
QCD describes the behavior of fundamental particles (quarks, gluons) that interact via the strong nuclear force.  At low temperatures, quarks and gluons are confined into bound states of mesons and baryons.  However, there is strong theoretical and experimental evidence that nuclear matter changes drastically when the temperature becomes large.  At sufficiently high temperature, quarks and gluons are liberated from the confines of mesons and baryons, forming a quark-gluon plasma (QGP).

\begin{figure}[b]
\begin{minipage}[c]{0.47\textwidth}
  \includegraphics[width=\textwidth]{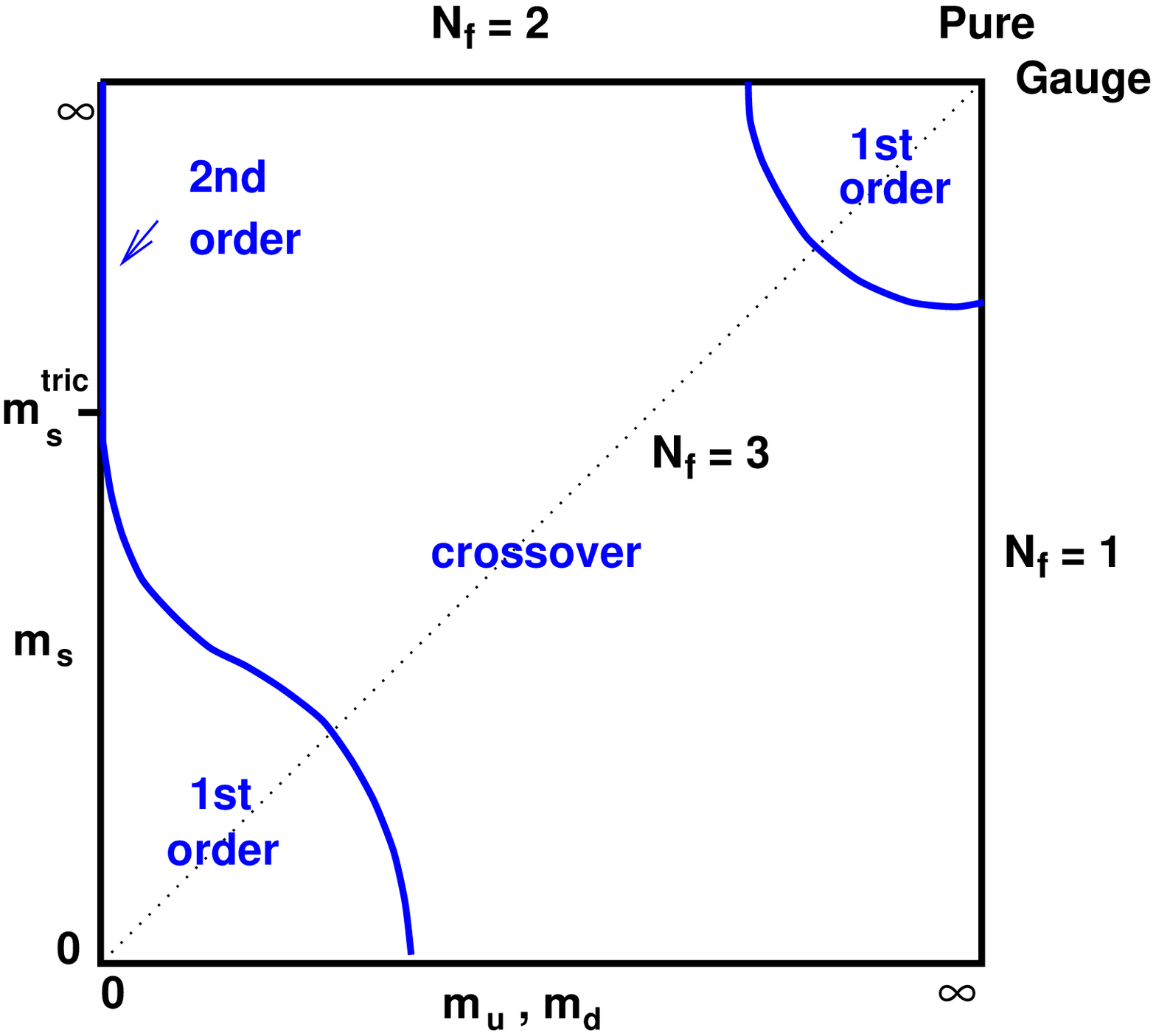}
  \caption{QCD phase diagram as a function of the light ($m_u, m_d$) and strange ($m_s$) quark masses.}
  \label{fig:phase_diagram}
\end{minipage}
\hspace{\fill}
\begin{minipage}[c]{0.47\textwidth}
  \includegraphics[width=\textwidth]{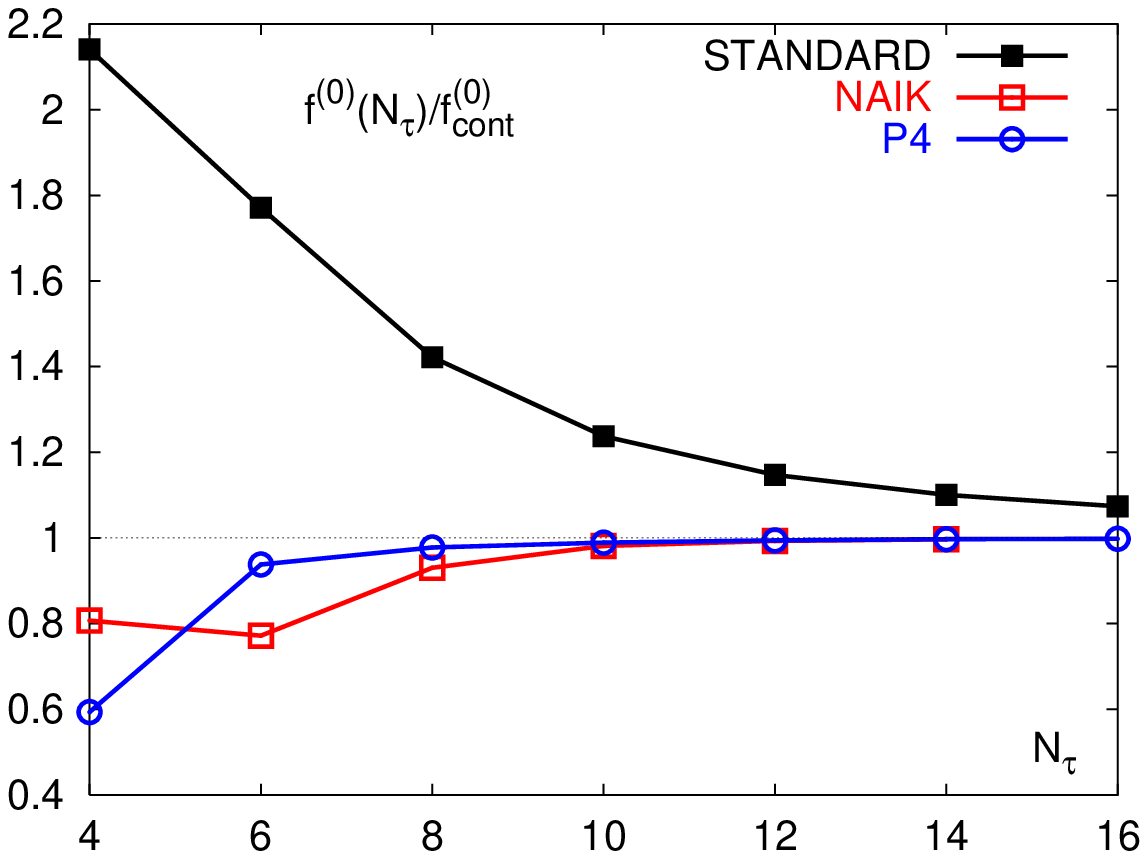}
  \caption{The free energy of an ideal gas on the lattice for naive staggered, p4, and Naik actions, normalized against the value in the continuum.}
  \label{fig:free_energy}
\end{minipage}
\end{figure}

We present here a study of high temperature QCD with 3 degenerate quark flavors.  Although 3f QCD is not directly applicable to the real world, we can still gain insight into QCD at high temperatures.  Figure \ref{fig:phase_diagram} shows the QCD phase diagram as a function of the quark masses.  For 3 flavors, it is believed that for small quark masses, we have a first-order chiral phase transition, which ends in a second-order critical point at $m_q=m_c$.  Conversely, with infinitely heavy quarks, we expect a deconfining phase transition.  For intermediate masses, the transition becomes a smooth crossover in a narrow temperature range\cite{Bernard:2004je, Karsch:2000kv}.

In this work we calculate the critical temperature $T_c$ and also attempt to explore the region where the crossover transition changes to first order.  By using finite-temperature lattices with $N_t=4$ and $N_t=6$, we can study the scaling properties of the p4 fermion action with two different choices of gauge-link smearing.  In our simulations, we use the inexact Hybrid-R algorithm, but we also make a comparison with the exact RHMC algorithm.  For more details, see our forthcoming paper\cite{Cheng:2006ab}.
\vspace{-5mm}

\section{Actions}
\vspace{-2mm}
For our calculation, we use the p4 staggered fermion action\cite{Heller:1999xz}.  The p4 action differs from the naive staggered action with the addition of a bent three-link term.  Thus, the p4 fermion matrix is written as:
\begin{displaymath}
M[U]_{ij} = \eta_i\left(c_1~\alink + c_{1,2}~\left[~\blinkbd ~+ \blinkbc ~+ \blinkba + ~\blinkbb~\right]\right)
\end{displaymath}
By choosing the coefficients appropriately, we can eliminate the violations to rotational symmetry up to $\mathit{O}(p^4)$ in the quark propogator.  As shown in Figure \ref{fig:free_energy}, the p4 action seems to have a better approach to the continuum limit for thermodynamic quantities, at least in the free-field case.

\begin{table}[h]
\begin{minipage}[c]{.32\textwidth}
\begin{center}
\begin{tabular}{|c|c|c|}
\hline
~ & \textbf{p4fat3} & \textbf{p4fat7}\\
\hline
$c_1$ & $\frac{3}{4}\frac{1}{1+6\omega}$ & -1/8\\
$c_3$ & $\frac{3}{4}\frac{\omega}{1+6\omega}$ & 1/16\\
$c_5$ & 0 & 1/64\\
$c_7$ & 0 & 1/384\\
$c_{1,2}$ & 1/48  & 1/48\\
\hline
\end{tabular}
\caption{Fermion action parameters.  $c_1$ is the one-link coefficient, while $c_{3,5,7}$ are for the 3, 5, and 7-link staples.  $c_{1,2}$ corresponds to the p4 term.  $\omega = 0.2$ for \textbf{p4fat3}}
\label{tab:fermion}
\end{center}
\end{minipage}
\hspace{\fill}
\begin{minipage}[l]{.65\textwidth}
In this 3f calculation, we use two different variants of the p4 action, differing only in the amount of gauge-link smearing.  The \textbf{p4fat3} action utilizes only the three-link staple in smearing the gauge fields.  We choose the weight for the three-link staple to be $\omega=0.2$.  This action has been used previously to study QCD thermodynamics\cite{Karsch:2000kv}.  The \textbf{p4fat7} action uses the three, five, and seven-link staples to more fully control the effects of taste-symmetry breaking.  The coefficients for \textbf{p4fat7} are chosen to eliminate the lowest order couplings between quarks and gluons that violate taste symmetry.  This method was also used to choose the parameters for the Asqtad action\cite{Orginos:1999cr}, with the exception that we omit the Lepage term, and do not use tadpole improvement.  Table \ref{tab:fermion} has a full list of the parameters.
\end{minipage}
\end{table}

For the gauge action, we employ a tree-level Symanzik action that includes the normal plaquette as well a planar 6-link rectangle term.

\begin{table}[bt]
\begin{center}
\begin{tabular}{|c|c|c|c|c|}
\hline
Action & $N_t$ & $N_\sigma$ & $m_q a$ & \# of $\beta$\\
\hline
p4fat3 & 4 & 8, 12, 16 & .005, .01, .025, .05, .1, .2 & 77\\
p4fat3 & 6 & 16 & .01, .02, .05, .1 & 49\\
\hline
p4fat7 & 4 & 8, 16, 32 & .01, .02, .035, .05, .1, .2 & 72\\
p4fat7 & 6 & 16, 32& .01, .02, .05, .2 & 61\\
\hline
\end{tabular}
\caption{Parameters for the finite-temperature simulations}
\label{tab:parameter}
\end{center}
\end{table}

\vspace{-2mm}
\section{Finite temperature simulations}
\vspace{-2mm}
We performed simulations for $N_t=4$ and $N_t=6$ with two different fermion actions at various different quark masses and for spatial volumes of $8^3, 12^3, 16^3$, and $32^3$ for $N_t=4$ and $16^3$ for $N_t=6$.  For each value of the quark mass, we simulated at several different values of the gauge coupling in order to sweep through the transition region.  All these simulations were done using the Hybrid-R algorithm\cite{Gottlieb:1987mq}, which suffers from being an inexact algorithm with \textit{O}$(\delta t^2)$ errors.  To minimize these errors, we choose a step size of $\delta t = 0.4 m_q$ for all our evolutions.  Table \ref{tab:parameter} gives a summary of the different simulation parameters.

In the chiral limit, the chiral condensate $\bar{\psi}\psi$ is a good order parameter for the chiral phase transition.  In the pure gauge theory, the Polyakov loop is a good signal for deconfinement.  In the intermediate regime, we do not expect a sharp phase transition, but both quantities should show the most rapid change in the transition region.

\begin{figure}[bt]
\vspace{-1mm}
  \begin{minipage}[c]{.47\textwidth}
    \includegraphics[width=\textwidth]{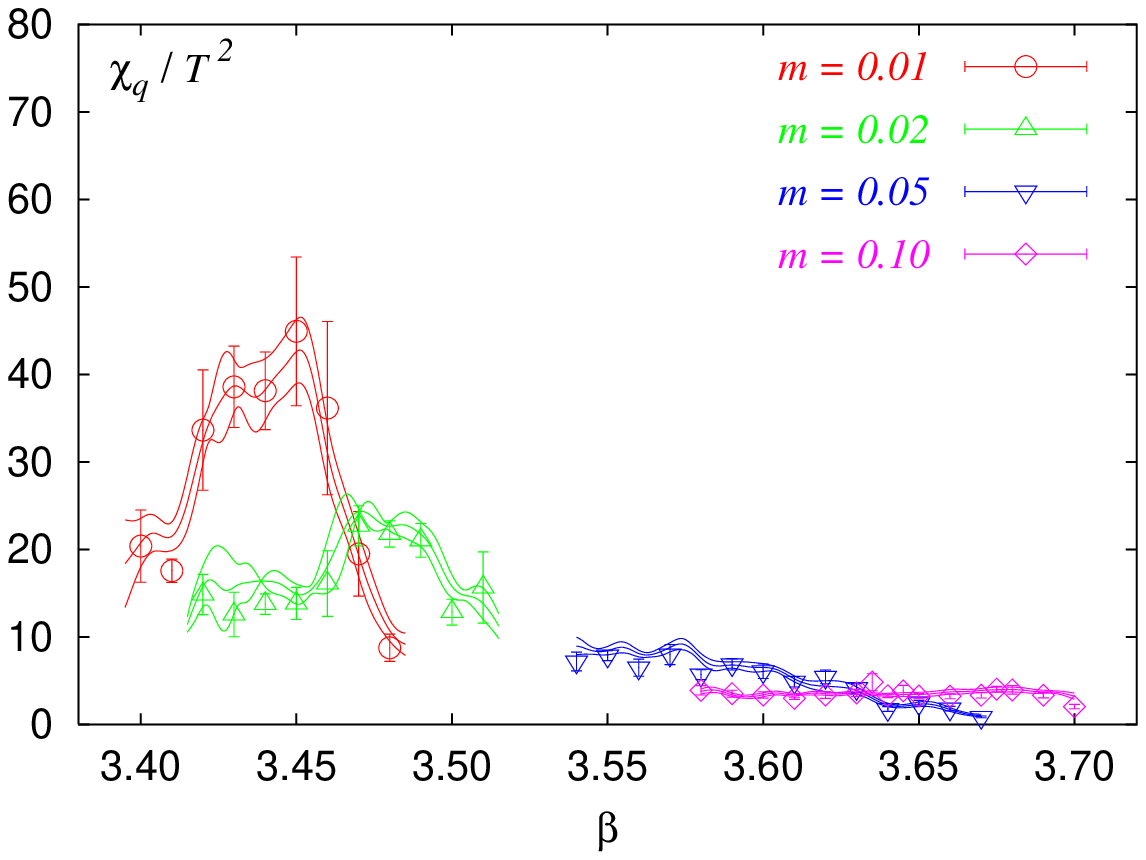}
    \caption{Chiral Condensate Susceptibility for p4fat3, $N_t=6$}
    \label{fig:p4fat3_chiral_sus}
  \end{minipage}
  \hspace{\fill}
  \begin{minipage}[c]{.47\textwidth}
    \includegraphics[width=\textwidth]{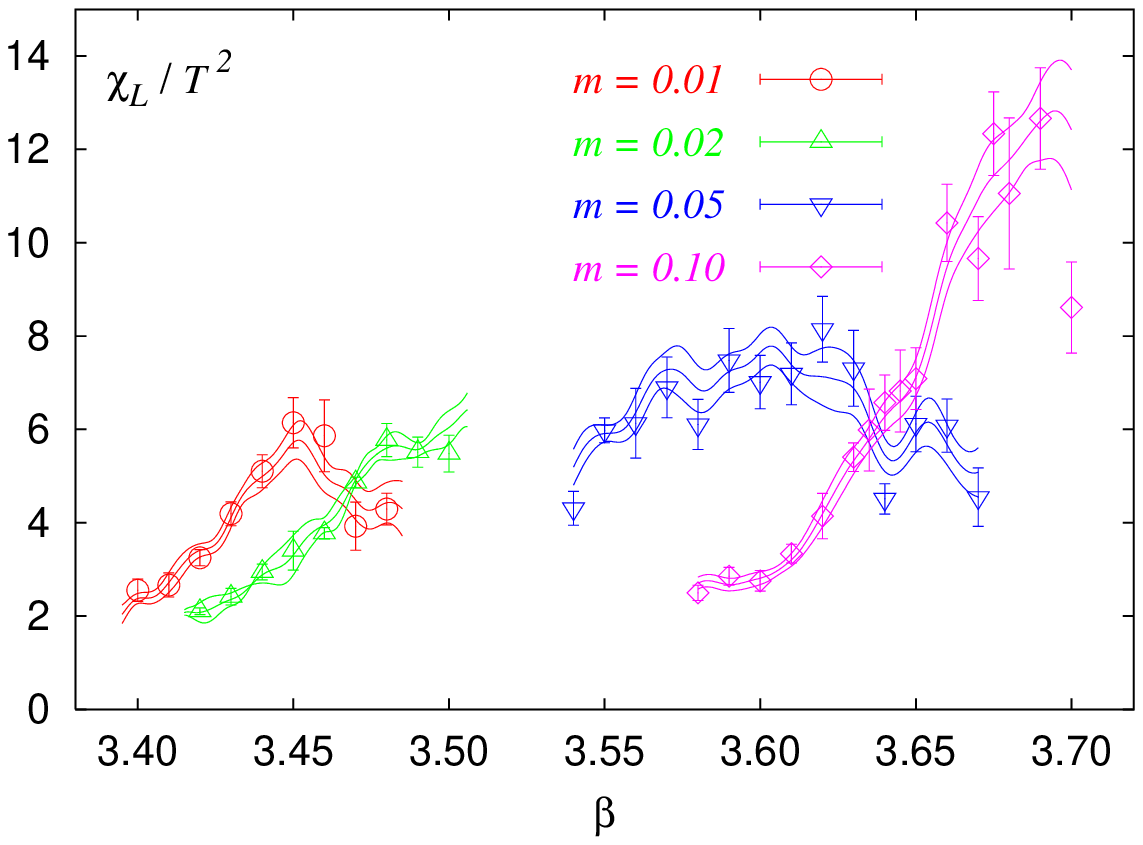}
    \caption{Polyakov Loop Susceptibility for p4fat3, $N_t=6$}
    \label{fig:p4fat3_polyakov_sus}
  \end{minipage}
\end{figure}

Accordingly, we measured the chiral condensate $\left<\bar{\psi}\psi\right>$, the Polyakov Loop $\left<L\right>$, and their susceptibilities ($\chi_q$, $\chi_L$)\symbolfootnote[2]{We measure only the disconnected part of the chiral susceptibility} in our finite temperature simulations.  The Polyakov loop is measured every trajectory, while $\bar{\psi}\psi$ is measured every ten trajectories, using at least ten different random sources per measurement.  We can then extract the critical value of the gauge coupling, $\beta_c$ by locating the peak in susceptibility for $\chi_q$ and $\chi_L$.  Figures \ref{fig:p4fat3_chiral_sus} and \ref{fig:p4fat3_polyakov_sus} show the chiral and Polyakov loop susceptibilities for \textbf{p4fat3} wtih $N_t=6$.  Table \ref{tab:betac} gives $\beta_c$ for different quark masses and volumes.

\begin{table}[b]
\begin{center}
\begin{tabular}{|c|c|c|c|c|c|}
\hline
$N_t$ & $m_q a$ & $N_{\sigma}$ & $\beta_{c,L}$ [from $\chi_L$] & $\beta_{c,q}$ [from $\chi_q$] & $\beta_c$ [averaged] \\
\hline
4 & 0.100 & 16 & 3.4800(27) & 3.4804(24)  & 3.4802(18)\\
~ & 0.050 & 16 & 3.3884(32) & 3.3862(47)  & 3.3877(34)   \\
~ &       & 8  & 3.4018(35) & 3.3930(201) & 3.4015(94) \\
~ & 0.025 & 8  & 3.3294(27) & 3.3270(28)  & 3.3283(31)   \\
~ & 0.010 & 16 & 3.2781(7)  & 3.2781(4)   & 3.2781(3)     \\
~ &       & 8  & 3.2858(71) & 3.2820(61)  & 3.2836(60) \\
~ & 0.005 & 16 & 3.2656(13) & 3.2678(12)  & 3.2667(24)  \\
~ &       & 12 & 3.2659(13) & 3.2653(12)  & 3.2656(10)  \\
\hline
6 & 0.200 & 16 & 3.8495(11) & 3.9015(279) & 3.8495(520) \\
~ & 0.100 & 16 & 3.6632(55) & 3.6855(105) & 3.6680(228) \\
~ & 0.050 & 16 & 3.6076(24) & 3.6189(328) & 3.6077(115) \\
~ & 0.020 & 16 & 3.4800(110) & 3.4800(80)  & 3.4800(65)  \\
~ & 0.010 & 16 & 3.4518(50) & 3.4510(83)  & 3.4516(44) \\
\hline
\end{tabular}
\caption{$\beta_c$ determined from the peaks in the Polyakov loop susceptibility and the chiral susceptibility using the \textbf{p4fat3} action.  The last column gives the average with combined statistical and systematic errors.}
\label{tab:betac}
\end{center}
\end{table}

\begin{figure}[bt]
  \vspace{-2mm}
  \begin{minipage}[c]{.47\textwidth}
    \includegraphics[width=\textwidth]{figs/chiral_3f_p4fat7.eps}
    \caption{Chiral Condensate for \textbf{p4fat7}, $N_t=4$}
    \label{fig:p4fat7_condensate}
  \end{minipage}
  \hspace{\fill}
  \begin{minipage}[c]{.47\textwidth}
    \includegraphics[width=\textwidth]{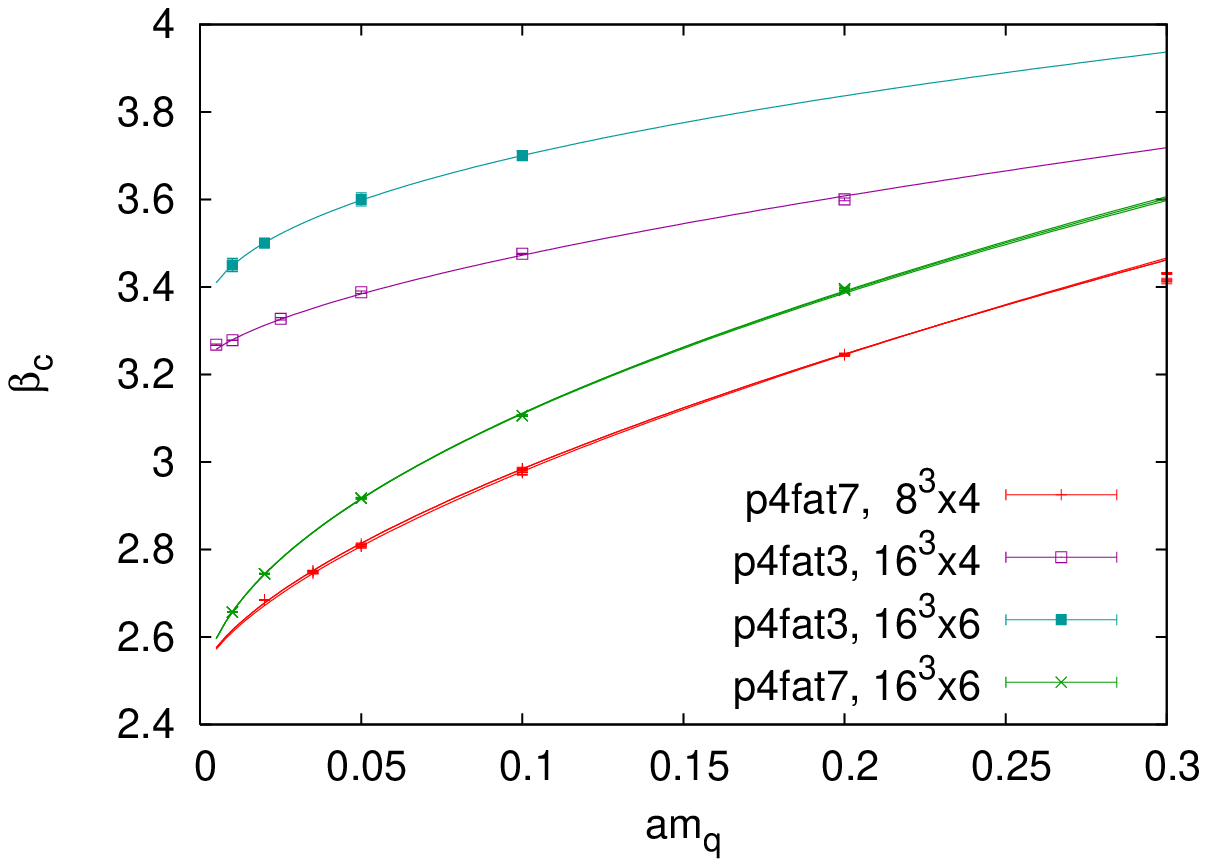}
    \caption{$\beta_c$ as a function of $m_q a$.  The top two lines are \textbf{p4fat3} and the bottom two are \textbf{p4fat7}}
    \label{fig:beta_c}
  \end{minipage}
\end{figure}

Figure \ref{fig:p4fat7_condensate} shows $\bar{\psi}\psi$ for the \textbf{p4fat7} action at $N_t=4$.  The sharpness of the transition for the chiral condensate and the Polyakov loop, as well as the volume scaling of the susceptibility peaks (not shown) seem indicative of a first order transition.  This first-order transition seems to persist even to very large values of the quark mass ($m_\pi \sim 800 MeV$), contrary to previous studies with staggered fermions\cite{Bernard:2004je, Karsch:2003va}.  Furthermore, as shown in Figure \ref{fig:beta_c}, the value of $\beta_c$ for $N_t=4$ and $N_t=6$ with \textbf{p4fat7} coincide as $m_q\rightarrow 0$.  We believe that the finite-temperature transition is being polluted by the presence of an unphysical, bulk phase transition.  This is supported by some exploratory studies at $N_t=8$.  As a result, we only use our studies with \textbf{p4fat3} to make statements about the 3f transition.  For more details on \textbf{p4fat7}, see Ref. \cite{Cheng:2006ab}

\begin{figure}[b]
  \begin{minipage}[c]{.47\textwidth}
    \includegraphics[width=\textwidth]{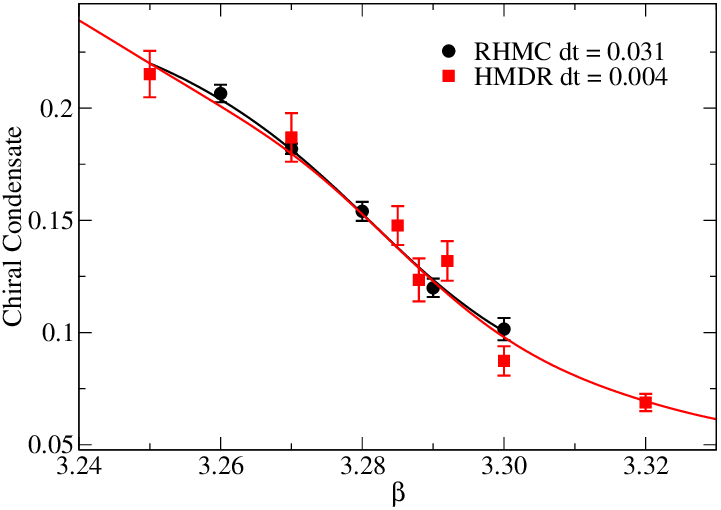}
    \caption{$\bar{\psi}\psi$ vs. $\beta$ for RHMC and R algorithms with \textbf{p4fat3}, $8^3\times 4, m_q=0.01$}
    \label{fig:p4fat3_rhmc}
  \end{minipage}
  \hspace{\fill}
  \begin{minipage}[c]{.47\textwidth}
    \includegraphics[width=\textwidth]{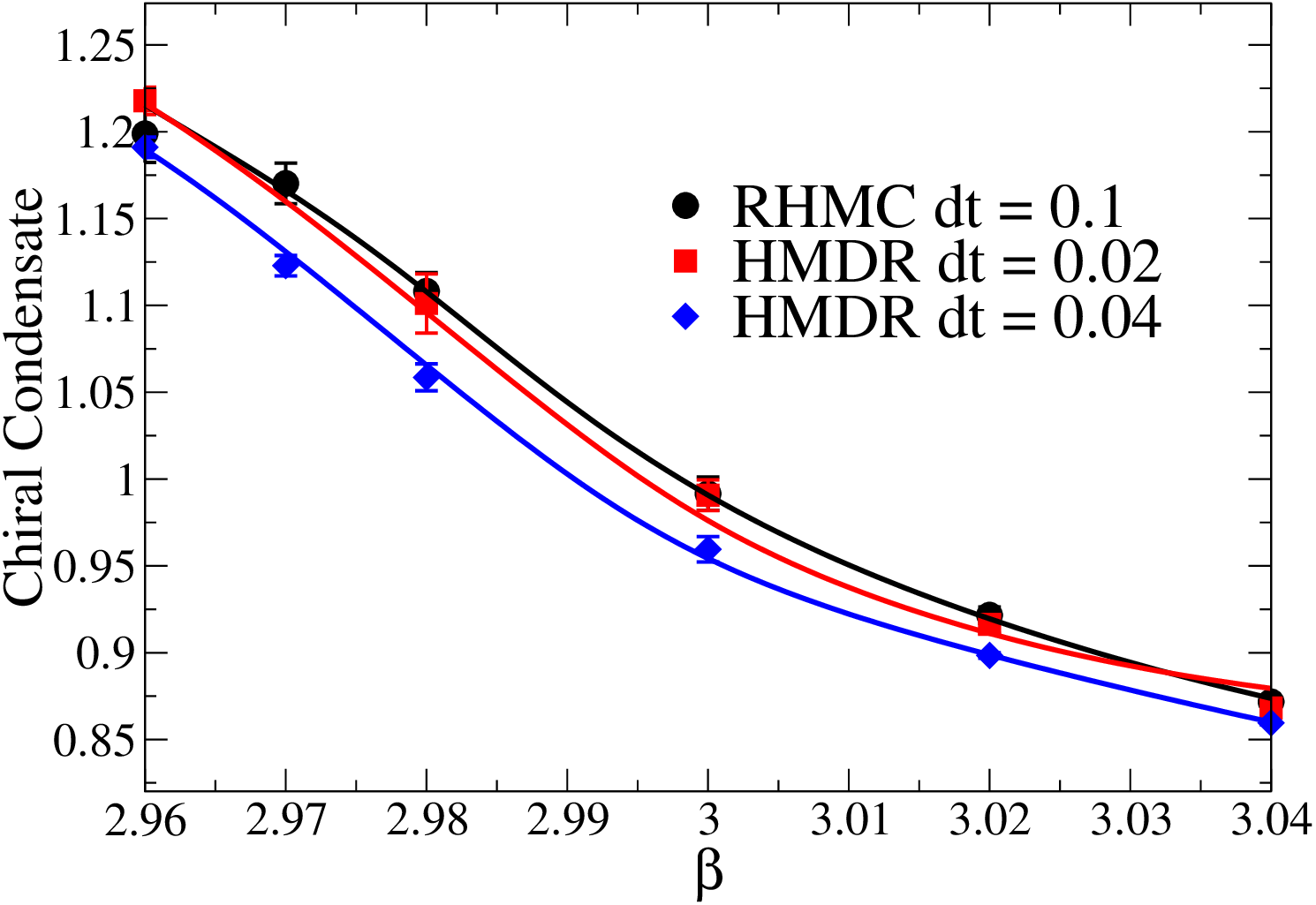}
    \caption{$\bar{\psi}\psi$ vs. $\beta$ for RHMC and R algorithms with \textbf{p4fat7}, $8^3\times 4, m_q=0.10$}
    \label{fig:p4fat7_rhmc}
  \end{minipage}
\end{figure}

\vspace{-3mm}
\section{Comparison of RHMC with R Algorithm}
\vspace{-2mm}
As our simulations were performed with the Hybrid-R algorithm, it is important to know to what extent the systematic errors in the R algorithm affect our results.  We have repeated some of our runs using the exact RHMC algorithm\cite{Clark:2004cp, Kennedy:1998cu}.  Figure \ref{fig:p4fat3_rhmc} is a comparison of the RHMC and R algorithms for \textbf{p4fat3}.  For this small quark mass ($m_q=0.01$), the step size we choose for the R algorithm is quite small ($\delta t = 0.004$).  At this step size, we have good agreement with RHMC.

Figure \ref{fig:p4fat7_rhmc} shows the same comparison for \textbf{p4fat7} at a larger quark mass ($m_q=0.10$).  For the step size used in our simulations ($\delta t = 0.04$), the errors in the R algorithm are apparent.  Only when we reduce the step size by a factor of two do we achieve agreement with the RHMC result.  However, in the calculation of $T_c$, the R algorithm seems mostly to shift only the bare parameters - the resulting systematic error in physical quantities seems to be less than the statistical error.

\begin{table}[b]
\begin{center}
\begin{tabular}{|c|c|c|c|c|c|c|c|}
\hline
$\beta$ &  $m_q a$ & \# traj & $m_{\pi} a$ & $m_{\pi 2} a$ & $m_{\rho} a$ & $r_0/a$ & $\sqrt{\sigma} a$\\
\hline     
3.3877 & 0.050 & 7800 & 0.7084(1) & 1.094(7) & 1.310(20)& 2.066(7)[7] & 0.552(12)[12]\\
3.3270 & 0.025 & 12000 & 0.5118(3) & 0.998(24) & 1.222(32) & 1.982(14)[13] & 0.564(11)[11] \\
3.2680  & 0.005 & 1500 &  0.2341(9) & 0.860(90) & 1.250(50) & 1.888(15)[9] & 0.587(17)[17]\\
\hline
3.46345 & 0.020 & 4420 & 0.4413(8) & 0.665(5) & 0.908(11) & 2.797(20)[20] & 0.404(6)[6]\\
3.4400 & 0.010 & 4290 & 0.3210(7) & 0.594(7) & 0.882(20) & 2.770(13)[13] & 0.405(6)[6]\\
\hline
\end{tabular}
\caption{Parameters for the zero temperature simulations, meson masses, the Sommer scale $r_0$, and the string tension $\sigma$.  The upper part of the table refers to scale setting runs for our $N_t=4$ lattices while the lower part to our $N_t=6$ calculations.}
\label{tab:zeroT}
\end{center}
\end{table}

\vspace{-2mm}
\section{Zero Temperature Scale Setting}
\vspace{-2mm}
In order to determine a physical scale for our lattices, we must measure zero temperature quantities such as the meson spectrum, the string tension $\sigma$, or the Sommer scale $r_0$.  Therefore, we also performed zero temperature simulations on $16^3\times 32$ volumes in the vicinity of the transition region.  On these lattices we measured both the meson spectrum and the static quark potential every tenth trajectory.

The static quark potential is extracted from the measurements of the Wilson loop.  The spatial transporters are constructed using iterative APE smearing.  Once the static quark potential has been determined, physical parameters such as the Sommer scale and the string tension can found by fitting to a three-parameter ansatz (Cornell potential):
\begin{displaymath}
V(r) = -\frac{\alpha}{r} + \sigma r + c
\end{displaymath}
where $\sigma$ is the string tension.  The Sommer scale $r_0$ is defined as:
\begin{displaymath}
r^2\frac{dV(r)}{dr}\vert_{r=r_0} ~~~= 1.65
\end{displaymath}
More details are found in Ref. \cite{Cheng:2006qk}

We also calculate the staggered meson spectrum, using point-wall quark propagators with a $\mathit{Z}_2$ wall source.  We obtain masses for the vector meson ($m_\rho$), the Goldstone pion ($m_\pi$), and the heavier non-Goldstone pion ($m_{\pi 2}$).  This allows us to check the scaling between $N_t=4$ and $N_t=6$, as well as to measure the extent of taste-symmetry breaking.  The parameters and results for the zero temperature measurements are summarized in Table \ref{tab:zeroT}.

\vspace{-3mm}
\section{Results and Conclusions}
\vspace{-3mm}
Our finite and zero temperature calculations allows us to express the critical temperature $T_c$ in terms of a physical scale set by the static quark potential - $T_c r_0$, for example.  We can then do a simultaneous chiral and continuum extrapolation to some appropriate physical limit:
\begin{displaymath}
T_c r_0(m_\pi,~N_t) = T_c r_0(m_{\pi c},~ N_t = \infty) + A\left((m_\pi r_0)^2 - (m_{\pi c} r_0)^2\right)^{1/\delta\beta} + B/N_t^2
\end{displaymath}
where $\delta\beta = 1.5654$ is the critical exponent for the $\mathit{Z}(2)$ universality class, corresponding to the endpoint in the first-order transition line.  This extrapolation is valid for $m_\pi > m_{\pi c}$ where the critical value of the pion mass, $m_{\pi c} \approx 0.15$, has been estimated in previous studies\cite{Karsch:2003va}.  In our case, all the points we have simulated are consistent with crossover behavior, with no indication of being in the first-order region.

This combined extrapolation results in a value $T_c r_0 = .439(8)[+1]$ at $m_\pi = m_{\pi c}$.  If we take $m_{\pi c} = 0$ and extrapolate to the chiral limit, we get $T_c r_0 = .419(9)$.  This is slightly smaller than the value obtained in our 2+1f calculation $T_c r_0 = 0.444(6)[+12][-6]$ in the chiral limit\cite{Cheng:2006qk}.
\begin{figure}[hbt]
\begin{minipage}[c]{.47\textwidth}
\includegraphics[width=\textwidth]{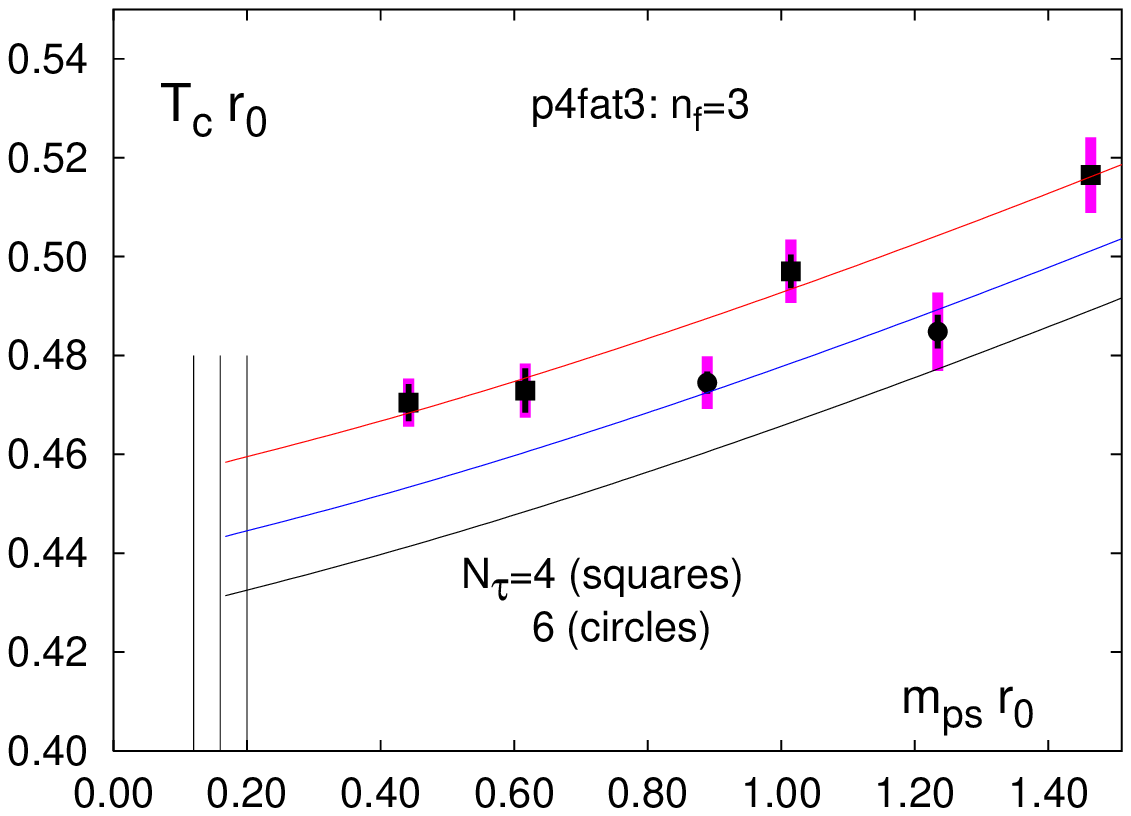}
\caption{Combined chiral, continuum extrapolation of $T_c r_0$.  Thin error bars are errors on $r_0/a$, thick error bars are for the determination of $\beta_c$.}
\end{minipage}
\hspace{\fill}
\begin{minipage}[c]{.47\textwidth}
\includegraphics[width=\textwidth]{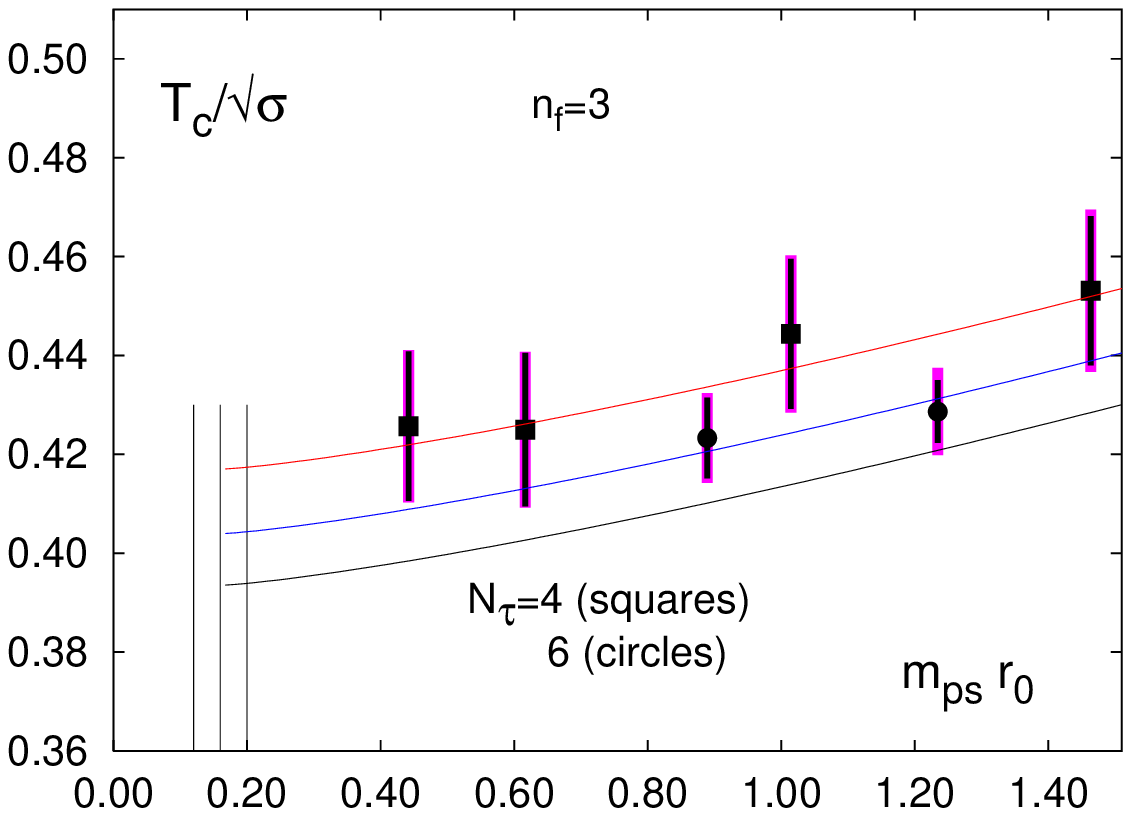}
\caption{Combined chiral, continuum extrapolation of $T_c/\sqrt{\sigma}$.  Thin error bars are errors on $\sqrt{\sigma} a$, thick error bars are for the determination of $\beta_c$.}
\end{minipage}
\end{figure}

\vspace{-5mm}
\section{Acknowledgments}
\vspace{-3mm}
This work has been supported in part by contracts DE-AC02-98CH1-886 
and DE-FG02-92ER40699 with the U.S. Department of Energy, 
the Helmholtz Gesellschaft under grant
VI-VH-041 and the Deutsche Forschungsgemeinschaft under grant GRK 881.  
Numerical simulations have been performed
on the QCDOC computer of the RIKEN-BNL research center, the DOE funded
QCDOC at BNL and the APE1000 at Bielefeld University.

\vspace{-2mm}
\bibliography{lat2006}

\begin{thebibliography}{10}
\expandafter\ifx\csname bibnamefont\endcsname\relax
  \def\bibnamefont#1{#1}\fi
\expandafter\ifx\csname bibfnamefont\endcsname\relax
  \def\bibfnamefont#1{#1}\fi
\expandafter\ifx\csname url\endcsname\relax
  \def\url#1{\texttt{#1}}\fi
\expandafter\ifx\csname urlprefix\endcsname\relax\def\urlprefix{URL }\fi
\expandafter\ifx\csname bibinfo\endcsname\relax \def\bibinfo#1#2{#2}\fi
\expandafter\ifx\csname eprint\endcsname\relax \def\eprint#1{#1}\fi

\bibitem{Bernard:2004je}
\bibinfo{author}{\bibfnamefont{C.}~\bibnamefont{Bernard}} \emph{et~al.}
  (\bibinfo{collaboration}{MILC}), \bibinfo{journal}{Phys. Rev.}
  \textbf{\bibinfo{volume}{D71}}, \bibinfo{pages}{034504}
  (\bibinfo{year}{2005}), \eprint{hep-lat/0405029}.

\bibitem{Karsch:2000kv}
\bibinfo{author}{\bibfnamefont{F.}~\bibnamefont{Karsch}},
  \bibinfo{author}{\bibfnamefont{E.}~\bibnamefont{Laermann}}, \bibnamefont{and}
  \bibinfo{author}{\bibfnamefont{A.}~\bibnamefont{Peikert}},
  \bibinfo{journal}{Nucl. Phys.} \textbf{\bibinfo{volume}{B605}},
  \bibinfo{pages}{579} (\bibinfo{year}{2001}), \eprint{hep-lat/0012023}.

\bibitem{Cheng:2006ab}
\bibinfo{author}{\bibfnamefont{M.}~\bibnamefont{Cheng}} \emph{et~al.}
  (\bibinfo{collaboration}{RBC-Bielefeld}), \bibinfo{journal}{in preparation}
  (\bibinfo{year}{2006}).

\bibitem{Heller:1999xz}
\bibinfo{author}{\bibfnamefont{U.~M.} \bibnamefont{Heller}},
  \bibinfo{author}{\bibfnamefont{F.}~\bibnamefont{Karsch}}, \bibnamefont{and}
  \bibinfo{author}{\bibfnamefont{B.}~\bibnamefont{Sturm}},
  \bibinfo{journal}{Phys. Rev.} \textbf{\bibinfo{volume}{D60}},
  \bibinfo{pages}{114502} (\bibinfo{year}{1999}), \eprint{hep-lat/9901010}.

\bibitem{Orginos:1999cr}
\bibinfo{author}{\bibfnamefont{K.}~\bibnamefont{Orginos}},
  \bibinfo{author}{\bibfnamefont{D.}~\bibnamefont{Toussaint}},
  \bibnamefont{and} \bibinfo{author}{\bibfnamefont{R.~L.} \bibnamefont{Sugar}}
  (\bibinfo{collaboration}{MILC}), \bibinfo{journal}{Phys. Rev.}
  \textbf{\bibinfo{volume}{D60}}, \bibinfo{pages}{054503}
  (\bibinfo{year}{1999}), \eprint{hep-lat/9903032}.

\bibitem{Gottlieb:1987mq}
\bibinfo{author}{\bibfnamefont{S.~A.} \bibnamefont{Gottlieb}},
  \bibinfo{author}{\bibfnamefont{W.}~\bibnamefont{Liu}},
  \bibinfo{author}{\bibfnamefont{D.}~\bibnamefont{Toussaint}},
  \bibinfo{author}{\bibfnamefont{R.~L.} \bibnamefont{Renken}},
  \bibnamefont{and} \bibinfo{author}{\bibfnamefont{R.~L.} \bibnamefont{Sugar}},
  \bibinfo{journal}{Phys. Rev.} \textbf{\bibinfo{volume}{D35}},
  \bibinfo{pages}{2531} (\bibinfo{year}{1987}).

\bibitem{Karsch:2003va}
\bibinfo{author}{\bibfnamefont{F.}~\bibnamefont{Karsch}} \emph{et~al.},
  \bibinfo{journal}{Nucl. Phys. Proc. Suppl.} \textbf{\bibinfo{volume}{129}},
  \bibinfo{pages}{614} (\bibinfo{year}{2004}), \eprint{hep-lat/0309116}.

\bibitem{Clark:2004cp}
\bibinfo{author}{\bibfnamefont{M.~A.} \bibnamefont{Clark}},
  \bibinfo{author}{\bibfnamefont{A.~D.} \bibnamefont{Kennedy}},
  \bibnamefont{and}
  \bibinfo{author}{\bibfnamefont{Z.}~\bibnamefont{Sroczynski}},
  \bibinfo{journal}{Nucl. Phys. Proc. Suppl.} \textbf{\bibinfo{volume}{140}},
  \bibinfo{pages}{835} (\bibinfo{year}{2005}), \eprint{hep-lat/0409133}.

\bibitem{Kennedy:1998cu}
\bibinfo{author}{\bibfnamefont{A.~D.} \bibnamefont{Kennedy}},
  \bibinfo{author}{\bibfnamefont{I.}~\bibnamefont{Horvath}}, \bibnamefont{and}
  \bibinfo{author}{\bibfnamefont{S.}~\bibnamefont{Sint}},
  \bibinfo{journal}{Nucl. Phys. Proc. Suppl.} \textbf{\bibinfo{volume}{73}},
  \bibinfo{pages}{834} (\bibinfo{year}{1999}), \eprint{hep-lat/9809092}.

\bibitem{Cheng:2006qk}
\bibinfo{author}{\bibfnamefont{M.}~\bibnamefont{Cheng}} \emph{et~al.}
  (\bibinfo{collaboration}{RBC-Bielefeld}), \bibinfo{journal}{Phys. Rev.}
  \textbf{\bibinfo{volume}{D74}}, \bibinfo{pages}{054507}
  (\bibinfo{year}{2006}), \eprint{hep-lat/0608013}.

\end{thebibliography}
\end{document}